\documentclass[5p, number, sort&compress]{elsarticle}

\usepackage[utf8]{inputenc}
\usepackage[T1]{fontenc}

\usepackage{hyperref}
\usepackage{xcolor}
\usepackage{amsmath}
\usepackage{siunitx}
\DeclareSIUnit\angstrom{\text{\AA}}
\usepackage{dcolumn}
\usepackage{xfrac}
\usepackage{microtype}
\usepackage{tabularx}
\usepackage{multirow}
\usepackage{booktabs}

\usepackage{graphicx}

\usepackage{lipsum}

\newcommand{\mytitle}{Role of defects in the thermodynamic stability
  of grain boundary phases at asymmetric tilt boundaries in copper}

\hypersetup{
  final,
  pdfborder={0 0 0},
  pdftitle={\mytitle},
  pdfauthor={Pemma, Langenohl, Saood, Choi, Janisch, Liebscher, Dehm, Brink},
  colorlinks=true,
  urlcolor=blue,
  linkcolor=blue,
  citecolor=blue
}

\begin{document}
\frenchspacing
\begin{frontmatter}
\title{\mytitle}

\author[1]{Swetha Pemma\fnref{fn1}}
\fntext[fn1]{Present address: \textit{Interdisciplinary Centre for Advanced
  Materials Simulation (ICAMS), Materials Informatics and Data
  Science, Ruhr-Universität Bochum, Universitätsstr. 150, 44801
  Bochum, Germany}}

\affiliation[1]{organization={Max Planck Institute for Sustainable Materials},
                addressline={Max-Planck-Stra\ss{}e 1},
                city={40237 D\"usseldorf},
                country={Germany}}

\author[1]{Lena Langenohl}

\author[1]{Saba Saood}

\author[1]{Yoonji Choi}

\author[2]{Rebecca Janisch}

\affiliation[2]{organization={Interdisciplinary Centre
                              for Advanced Materials Simulation
                              (ICAMS), Ruhr-Universität Bochum},
                city={44780 Bochum},
                country={Germany}}

\author[1,3,4]{Christian H. Liebscher}    %

\affiliation[3]{organization={Research Center Future Energy
                              Materials and Systems,
                              Ruhr-Universit\"at Bochum},
                city={44801 Bochum},
                country={Germany}}
\affiliation[4]{organization={Faculty of Physics and Astronomy,
                              Ruhr-Universit\"at Bochum},
                city={44801 Bochum},
                country={Germany}}

\author[1]{Gerhard Dehm\corref{cor1}}
\ead{dehm@mpi-susmat.de}

\author[1]{Tobias Brink\corref{cor1}}
\ead{t.brink@mpi-susmat.de}
\cortext[cor1]{Corresponding author}

\date{\today}

\begin{abstract}
  Grain boundaries can exist as different grain boundary phases (also
  called complexions) with individual atomic structures.  The
  thermodynamics of these defect phases in high-angle grain boundaries
  were studied mostly with atomistic and phase field computer
  simulations, but almost exclusively for special, symmetric
  boundaries.  Here, we use molecular dynamics simulations combined
  with structure search methods, as well as scanning transmission
  electron microscopy experiments to take a step towards understanding
  more general grain boundaries. Using the example of $\Sigma$37c
  $[11\overline{1}]$ tilt boundaries in Cu, we show how the grain
  boundary phase transition on a symmetric boundary plane is changed
  by the geometrically necessary defects introduced in inclined,
  asymmetric boundaries.  We analyze the disconnections---which are
  dislocation-like line defects of grain boundaries---both in the
  simulations, as well as in experimental Cu and Al samples.  A main
  finding is that defect energies can have a major influence on the
  stability of grain boundary phases, even at small
  inclinations. Furthermore, some defects are not able to effect large
  inclinations. At that point, defective asymmetric GB phases compete
  with grain boundaries faceting into the adjacent symmetric GB phases.
\end{abstract}


\journal{arXiv} 

\end{frontmatter}

\newcounter{supplfigctr}
\renewcommand{\thesupplfigctr}{S\arabic{supplfigctr}}
\newcounter{supplsecctr}
\renewcommand{\thesupplsecctr}{\Roman{supplsecctr}}
{\refstepcounter{supplsecctr}\label{sec:suppl:phase-fractions}}
{\refstepcounter{supplfigctr}\label{fig:suppl:phase-fractions}}
{\refstepcounter{supplsecctr}\label{sec:suppl:dichrom-S37}}
{\refstepcounter{supplfigctr}\label{fig:suppl:dichrom-S37}}
{\refstepcounter{supplsecctr}\label{sec:suppl:cryst-poss-def}}
{\refstepcounter{supplsecctr}\label{sec:suppl:combined-defect-energies}}
{\refstepcounter{supplfigctr}\label{fig:suppl:combined-defect-energies}}
{\refstepcounter{supplsecctr}\label{sec:suppl:phase-junc}}
{\refstepcounter{supplfigctr}\label{fig:suppl:phase-junc}}
{\refstepcounter{supplsecctr}\label{sec:suppl:stem}}
{\refstepcounter{supplfigctr}\label{fig:suppl:Cu_exptimages}}
{\refstepcounter{supplfigctr}\label{fig:suppl:Al_exptimages}}
{\refstepcounter{supplsecctr}\label{sec:suppl:domino-terrace-stress}}
{\refstepcounter{supplfigctr}\label{fig:suppl:domino-terrace-stress}}
{\refstepcounter{supplsecctr}\label{sec:suppl:facet-fractions}}
{\refstepcounter{supplfigctr}\label{fig:suppl:facet-fractions}}

\section{Introduction}

Grain boundaries (GBs) are metastable planar defects that occur in all
crystalline materials, depending on the synthesis route. They strongly
influence mechanical properties \cite{Hall1951, Petch1953,
  lejcek2010grain, priester2012, Han2018}, diffusion \cite{Kaur1995,
  lejcek2010grain, Divinski2012, Divinski2013}, electrical
conductivity \cite{Taylor1952, Andrews1969, Mayadas1970, Lormand1982,
  Bishara2021}, thermal conductivity \cite{Klemens1994, Goel2016a,
  Hickman2020, Isotta2023, Isotta2024}, and others. It has been
established theoretically \cite{GibbsVol1, Hart1968, Cahn1979,
  Cahn1982, Rottman1988, Frolov2012I, Frolov2012II, Han2017} that GBs
can occur in different GB phases \cite{Frolov2015}, also called
complexions \cite{Tang2006, Dillon2007, Cantwell2014, Cantwell2020},
or more generally defect phases \cite{Korte-Kerzel2022, Tehranchi2024,
  Zhou2025}. A thermodynamic treatment of these metastable GB phases
as defects of stable bulk phases is possible by presupposing the
existence of the GB. The GB phase with the lowest excess free energy
is then the (meta-)stable interface phase at the given thermodynamic
conditions. Recently, the formalism has been extended to the
hypothetical case of a stable interface phase with zero excess free
energy \cite{Li2025}.  Analogous to bulk phases, GB phase transitions
can be due to changes of composition \cite{Frolov2015seg, Luo2020a,
  Cantwell2020, Futazuka2022, BuenoVilloro2023a, Devulapalli2024}, but
they can even exist in single-element materials, where just the GB
structure changes \cite{MILLS1992, Mills1993, Frolov2013, Hickman2017,
  Aramfard2018, Zhu2018, Frolov2018bcc, Frolov2018StructuresAT,
  Yang2020, Meiners2020, Langenohl2022, Winter2022, Brink_2023,
  Chen2024, Choi2025}.  %

GB phases are important, because they affect how the GB impacts the
macroscopic material properties.  This is true even in single-element
systems, where for example GB mobility \cite{Frolov2014, pemma2024} or
diffusion \cite{Divinski2012, Divinski2013} are affected by GB phases.

GBs are characterized by five macroscopic parameters, covering the
misorientation axis and angle $\theta$ between the abutting
crystallites and the crystallographic planes that terminate on both
sides of the GB \cite{lejcek2010grain, priester2012}. Here, we cover
single-element systems with high-angle tilt GBs, where the rotation
axis between the grains lies within the GB plane.  Often, GBs with
crystallographically equivalent planes on both sides of the GB are
called symmetric. Finer distinctions exist, where GBs that are mirror
planes are called symmetric and those with only equivalent planes, but
no mirror symmetry, are called quasi-symmetric \cite{Morawiec2012}. We
will generally use the term ``symmetric'' to refer to both the
quasi-symmetric and the symmetric case for simplicity.  Most studies
of GB thermodynamics have exclusivley considered special GBs, such as
symmetric GBs \cite{Frolov2013, Hickman2017, Aramfard2018, Zhu2018,
  Frolov2018bcc, Frolov2018StructuresAT, Yang2020, Winter2022,
  Brink_2023, Chen2024}, which typically have the lowest energy.
However, tilt GBs found in real materials are never perfect and
deviate from the ideal case by having a twist component or they are
inclined by an angle $\phi$ from the nearest symmetric GB plane
(asymmetric tilt GBs). GB phase transitions without segregation in
these common cases are less well studied, although similar GB phases
have been observed in symmetric and asymmetric GBs of pure Cu
\cite{Meiners2020, Langenohl2022}.

There are two ways that the atomic structures can accommodate an
inclination away from the symmetric plane. First, the GB can facet
into the two nearest symmetric facets, increasing its area $A$, but
potentially decreasing its overall excess free energy $\gamma$
\cite{Wagner1974, Brokman1981, Vitek1983b, Dahmen1991, Straumal2001,
  Brown2007, Tschopp2007, Tschopp2007b, Banadaki2016, Abdeljawad2016,
  Medlin2017, Dobrovolski2024, Ding2025}. Alternatively, defects with
step character can lead to inclination \cite{Vitek1983b, Tschopp2007,
  Trautt_2012, Race2015, Hadian2016, Meiners2020a, Saba2023,
  Ding2025}.
Apart from pure steps, common line defects in GBs are disconnections,
which are characterized by a mode (\textbf{b}, \textit{h})
\cite{Han2018}. The mode is a specific combination of the dislocation
character (Burgers vector \textbf{b}) of the defect and a
corresponding step height $h$. Disconnections are responsible for GB
plasticity in the form of shear-coupled motion \cite{Han2018} and can
also influence macroscopic properties, such as diffusion
\cite{Paul2014, Sevlikar2024}. In terms of deviations from special
GBs, the \textbf{b} component can change the twist and tilt
misorientation and the \textit{h} component can influence the
inclination of the GB. Depending on the mode, disconnections can have
different core and elastic energies \cite{Han2018}. The core energies,
in turn, depend on the GB phase \cite{pemma2024}.  Thus, we expect
that geometrically necessary disconnections and steps in asymmetric
tilt GBs also influence the thermodynamic stability of the GB phases.

\begin{figure}
    \centering
    \includegraphics[width=\linewidth]{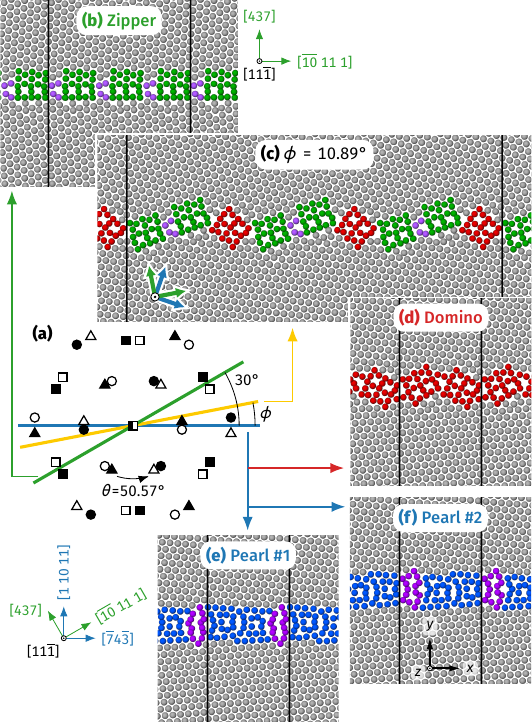}
    \caption{Bicrystallography of $\Sigma37$c $[11\overline{1}]$ tilt
      GBs in Cu. (a) Part of the dichromatic pattern. The blue line
      represents the quasi-symmetric $\{1~10~11\}$ GB planes, while
      the green line represents the symmetric $\{347\}$ GB
      planes. Asymmetric GBs, with inclinations $\phi$ between these
      two cases are the subject of the present study (yellow
      line). The coordinate system in the bottom left represents the
      corresponding crystal directions in the upper crystallite. (b)
      The zipper GB phase at $\phi=\ang{30}$ depicted with the crystal
      orientations as indicated by the coordinate axes to the right of
      the image. The vertical black lines represent the periodic unit
      cell of the GB structure. (c) An asymmetric GB at intermediate
      $\ang{0} < \phi < \ang{30}$ obtained as a mix of zipper (b) and
      domino (d) structures. (d)--(f) The domino and pearl phases on
      the $\{1~10~11\}$ GB planes. Note the difference in the purple
      structure for pearl \#1 and \#2. An asymmetric pearl phase is
      not depicted and will be studied in this work.}
    \label{fig:bicrystallography}
\end{figure}

Here, we study asymmetric $\Sigma37$c $[11\overline{1}]$ tilt GBs in
Cu, because their symmetric variant exhibits a GB phase transition
between two experimentally-observed GB phases with changing
temperature \cite{Langenohl2022}.  We investigated the GBs using
exhaustive structure search, molecular dynamics (MD) simulations, and
scanning transmission electron miscroscopy (STEM) of Cu and Al
samples.  Our goal is to determine how the inclination affects the
atomic structure and how defects and faceting in turn affect the
thermodynamics of the GB phases.

\section{Bicrystallography}

First, we will revisit the previously-studied GB phases and
crystallography (Fig.~\ref{fig:bicrystallography}). The \{1 10
11\}-type GBs (repeated every \ang{60} due to the trifold symmetry
along $[11\overline{1}]$) are quasi-symmetric GBs in the nomenclature
of Ref.~\cite{Morawiec2012}, in that they exhibit equivalent
crystallographic planes on both sides of the GB, but the abutting
crystallites are not mirrors of each other and are instead connected
by another symmetry operation \cite{Morawiec2012}. We arbitrarily
define this plane as having inclination $\phi = \ang{0}$. This plane
can have two types of GB phases, namely the ``domino''
(Fig.~\ref{fig:bicrystallography}(d)) and the ``pearl''
(Fig.~\ref{fig:bicrystallography}(e)--(f)) GB phase
\cite{Langenohl2022}. The pearl phase has two microstates with
slightly different structures and energies. The domino structure is
stable at low temperatures, while the pearl phase is stable at higher
temperatures. The GB phases can transform into each other without
requiring diffusion \cite{Langenohl2022}.  At $\phi = \ang{30}$, we
find the symmetric \{347\} GB planes. These GB planes are truly
symmetric mirror planes and exhibit only the zipper GB phase
\cite{Langenohl2023, Saba2024}
(Fig.~\ref{fig:bicrystallography}(b)). We will from now on no longer
make the distinction between quasi-symmetric and symmetric GBs, and
call both symmetric for simplicitly. A noteworthy feature of the
domino phase is that it can also be regarded as consisting of
nanoscaled facets of the zipper structure, which are driven
energetically to their minimal length \cite{brink2024}. This allows
the easy construction of asymmetric domino/zipper GBs by simply
elongating one facet of the domino structure as shown in
Fig.~\ref{fig:bicrystallography}(c). We will therefore call the set of
domino, zipper, and corresponding asymmetric GB phases simply the
domino/zipper phase. The asymmetric pearl phase and the relative
thermodynamic stability of these asymmetric GB phases have not been
studied, yet, and are therefore the topic of the present work.

\section{Methods}

\subsection{General simulation parameters}

To model Cu, we used the Mishin et al.\ embedded atom method (EAM)
potential \cite{Mishin2001} with the simulation code LAMMPS
\cite{Plimpton1995, Thompson_2022}. For later use, we determined the
lattice constant at $T = 0$ with this potential as
\begin{equation}
  a = \SI{3.615}{\angstrom}
\end{equation}
and calculated the values at a range of finite temperatures using the
protocol described in Ref.~\cite{Freitas2016}.

\subsection{Asymmetric grain boundary structures and
  structure search}

We employed different strategies to find asymmetric GB structures in
the present work. As a convention, we oriented the GBs in our
simulations and (experimental) images such that the $z$ direction
coincides with the $[11\overline{1}]$ tilt axis and the $y$ direction
is the GB normal. The $x$ direction is then the orthogonal direction
within the GB plane. Sometimes, where noted, the $x$ and $y$
directions instead correspond to the closest symmetric GB plane.

First, we constructed a range of domino/zipper asymmetric GBs by
starting from the domino structure and increasingly elongating one of
the nanofacets \cite{brink2024}, see also
Fig.~\ref{fig:bicrystallography}(c). For the bordering symmetric cases we
used the following crystal directions for the upper (``top'') and
lower (``bot'') crystallite:
\begin{align}
  \mathbf{x}^\text{top}_{\ang{0}} &= [\overline{7}4\overline{3}] &
  \mathbf{y}^\text{top}_{\ang{0}} &= [1~10~11] &
  \mathbf{z} &= [11\overline{1}]
  \\
  \mathbf{x}^\text{bot}_{\ang{0}} &= [\overline{4}73] &
  \mathbf{y}^\text{bot}_{\ang{0}} &= [10~1~11]
  \\
  \mathbf{x}^\text{top}_{\ang{30}} &= [\overline{10}~11~1] &
  \mathbf{y}^\text{top}_{\ang{30}} &= [437]
  \\
  \mathbf{x}^\text{bot}_{\ang{30}} &= [\overline{1}~11~10] &
  \mathbf{y}^\text{bot}_{\ang{30}} &= [7\overline{3}4].
\end{align}
The construction by elongation of a nanofacet and subsequent rotation
yielded various asymmetric GBs that are periodic in $x$ and $z$
direction.

Using the resulting crystal directions and periodicities, we also
performed an exhaustive structure search with the software GRIP
\cite{Chen2024}. This code searches for GB structures not only by
varying relative displacements between the upper and lower crystal
($\gamma$-surface method) and running short MD simulations, but also
by adding or removing atoms in the GB region, yielding GB structures
that cannot be found with a constant number of atoms \cite{Frolov2013,
  Hickman2017, Zhu2018, Chen2024}. Due to the large size of the
periodic cell of the asymmetric GBs, we did not run the search for any
supercells, but only the smallest possible sizes in the periodic $x$
and $z$ directions.

For all structures, we fixed the lattice constant to its
zero-temperature value $a$, so that the GBs were constrained in $x$
and $z$ directions by the bulk material, only allowing volume
relaxation normal to the GB \cite{Frolov2012I, Frolov2012II}. We
therewith obtained the ground-state GB energy
\begin{equation}
  \label{eq:gamma0}
  \gamma_0 = [E_\text{pot}],
\end{equation}
where $E_\text{pot}$ is the system's potential energy at $T = 0$ and
the square brackets mark the excess of any extensive property $Z$
\cite{Cahn1979, Frolov2012II}, defined as
\begin{equation}
  \label{eq:excess}
  [Z] = \frac{Z_\text{GB} - \frac{N_\text{GB}}{N_\text{bulk}}Z_\text{bulk}}{A}.
\end{equation}
Here, $Z_\text{GB}$ is calculated in a region containing a GB with
area $A$ and $N_\text{GB}$ atoms, while the reference state is a
defect free crystal region with the property $Z_\text{bulk}$ for
$N_\text{bulk}$ atoms.

\subsection{Annealing with molecular dynamics}

Additionally, we performed MD simulations to anneal samples at various
temperatures in order to determine their equilibrium GB phases. For
these simulations we used a time integration step of \SI{2}{fs} and a
Nos\'e--Hoover thermostat at the desired temperature $T$. We again
employed periodic boundary conditions in $x$ and $z$ direction, as
well as open boundaries in $y$ direction. In the periodic directions,
the size of the simulation cell was set in accordance with the
finite-temperature lattice constant of the bulk material. In $x$
direction, the various symmetric and asymmetric GBs had different
periodic lengths. In order to keep the simulations of roughly
comparable size (on the order of 150\,000 to 170\,000 atoms), we
repeated the periodic segments as necessary to reach sizes between
\SI{700}{\angstrom} and \SI{800}{\angstrom}. In the open $y$
direction, we kept a size of \SI{400}{\angstrom} to minimize
interaction of the GBs' stress fields with the surface. In $z$
direction, along the tilt axis, we only used a single unit cell with
length $a \sqrt{3} \approx \SI{6.3}{\angstrom}$. While this also
minimizes the required computational power, we mainly chose this value
to accelerate the kinetics of the GB phase transition at temperatures
close to room temperature. It was shown before that the GB phase
junction between two GB phases moves slower the larger the system size
along the tilt axis \cite{Meiners2020}. The periodic lengths were
scaled and fixed in accordance with the bulk material's lattice
constant at the given temperature $T$. Annealing simulations were then
performed for \SI{100}{ns}. Subsequently, the structures were rescaled
to the zero-temperature lattice constant $a$ and minimized to extract
the ground-state GB energy $\gamma_0$ for comparison with the results
of the structure search.

\subsection{Quasi-harmonic approximation}

The stability of the GB phases at finite temperatures is defined by
the excess free energy
\begin{equation}
  \label{eq:gamma}
  \gamma = [G] = [U] - T[S]
\end{equation}
at zero externally applied stress \cite{Frolov2012II}. While the
excess internal energy $[U]$ can be obtained directly from MD
simulations, the excess entropy $[S]$ requires other methods.

Here, we use the quasi-harmonic approximation (QHA) for the
vibrational part of the free energy \cite{Foiles1994,
  Freitas2018}. For this, we computed the force constant matrix with
the \texttt{dynamical\textunderscore{}matrix} command in LAMMPS, which
was diagonalized to obtain the eigenfrequencies $\nu_i$. We considered
only a classical approximation for the excess free energy since (i) we
previously found the difference to the quantum mechanical
approximation to be minor above \SI{50}{K} \cite{Langenohl2022,
  Choi2025} and (ii) we compare the results to classical MD
simulations. The free energy of a system is thus calculated as
\begin{equation}
  \label{eq:free-energy-qha}
  F(T) = E_\text{pot}(T) + k_BT \sum_{i=1}^{3N-3} \ln\frac{h\nu_i}{k_BT}
\end{equation}
and $G$ is obtained by calculating $F$ at the equilibrium lattice
constant at each temperature \cite{Foiles1994, Freitas2018}.

For asymmetric GBs, the computation of the excess properties from the
QHA data is more complex. The free energy from the QHA must always be
calculated for the whole system \cite{Langenohl2022} and thus requires
the subtraction of the surface energies $\Gamma$, which differ for the
upper crystal ($\Gamma_\text{top}$) and lower crystal
($\Gamma_\text{bot}$). For each GB, we thus computed the free energies
for three slabs with open surfaces in $y$ direction and periodic
boundaries in $x$ and $z$: $G_\text{GB}$ for a slab with a GB, as well
as $G_\text{top}$ and $G_\text{bot}$ for defect-free slabs
corresponding to the orientation and surfaces of the upper and lower
crystallite. Additionally, we calculated $G_\text{fcc}$ for a fully
periodic, defect-free fcc crystal. Thus, we compute
\begin{align}
  \Gamma_\text{top} &= \frac{G_\text{top} - N_\text{top}\frac{G_\text{fcc}}{N_\text{fcc}}}{2A}\\
  \Gamma_\text{bot} &= \frac{G_\text{bot} - N_\text{bot}\frac{G_\text{fcc}}{N_\text{fcc}}}{2A}\\
  \label{eq:gamma-asymm-1}
  \gamma &= \frac{G_\text{GB} - N_\text{GB}\frac{G_\text{fcc}}{N_\text{fcc}}}{A}
           - \Gamma_\text{top} - \Gamma_\text{bot},
\end{align}
where the GB area $A$ is equal to the surface area on one side of the
slab due to our construction of the simulation cells. Alternatively,
we can also write
\begin{align}
  \label{eq:gamma-asymm-2}
  \gamma &= \frac{G_\text{GB}
                  - \frac{G_\text{top}}{2}
                  - \frac{G_\text{bot}}{2}}
                 {A}
\end{align}
if $N_\text{GB} = N_\text{top} = N_\text{bot}$. This is useful to
determine if the system containing the GB is big enough to avoid
interactions between surface and GB. If the system is too small, the
error can be estimated by the discrepancy between
Eqs.~\ref{eq:gamma-asymm-1} and \ref{eq:gamma-asymm-2}.

\subsection{Experimental methods}

We additionally investigated several asymmetric GBs experimentally
with STEM. For this, we used Cu and Al samples, to be able to compare
the results of two different fcc metals. In order to obtain
high-resolution atomic images of tilt GBs, $\{ 111\}$-textured films
are required. The recipe and details of the growth of Cu thin films
with a $\{ 111 \}$ surface orientation can be found in
Refs.~\cite{Langenohl2022, Langenohl2023}, as the same thin film was
used for the present experiments. Similarly, $\{ 111 \}$-textured
Al thin films were deposited following the procedure described
in Ref.~\cite{Saba2023}. The resulting films in both Cu and Al possess
grains with sizes of tens of micrometers with a sharp
$\{ 111 \}$ texture separated by $\langle 111 \rangle$ tilt GBs.

For selecting and identifying $\Sigma$37c GBs in Cu, the film was
scanned with a Thermo Fisher Scientific Scios2HiVac dual-beam
secondary electron microscope (SEM) equipped with an
electron-backscattered diffraction (EBSD) detector. The focused Ga$^+$
ion beam (FIB) of the dual-beam SEM was used to lift out the GBs of
interest. Thinning of the lamellae was performed with ion beam voltage
and current being gradually reduced from their starting values of
\SI{30}{kV} and \SI{0.1}{nA} down to \SI{5}{kV} and \SI{16}{pA}.

For Al, the thin films were analyzed by EBSD at \SI{20}{kV} using a
JEOL JSM-6490 SEM. Given that Ga implantation can lead to
embrittlement in GBs \cite{Nam2009}, a Xe-based plasma FIB (PFIB)
milling process was employed for the preparation of the Al
lamellae. The PFIB milling was carried out using a Helios G3 Cx
dual-beam SEM/FIB system (Thermo Fisher Scientific) and the parameters
applied for plan-view STEM sample preparation are detailed in
Ref.~\cite{Saba2023}.

All samples were subsequently analyzed in a probe-corrected FEI Titan
Themis 80-300 (Thermo Fisher Scientific). A high-brightness field
emission gun at a voltage of \SI{300}{kV} and currents of 70 to
\SI{80}{pA} were used to scan the area of interest. The STEM images
were recorded with a high-angle annular dark-field (HAADF) detector
(Fishione Instruments Model 3000) with collection angles of 78 to
\SI{200}{mrad} and a semiconvergence angle of \SI{17}{mrad}. The
images shown in this study are averaged over 50 to 100 frames and
filtered with a background substraction filter, Butterworth filter,
and Gaussian filter to reveal the atomic structures of the GBs,
ensuring that the structures of the original image were not modified
by using the applied filters.

\subsection{Determination of Burgers vectors}

\begin{figure}
    \centering
    \includegraphics[width=\linewidth]{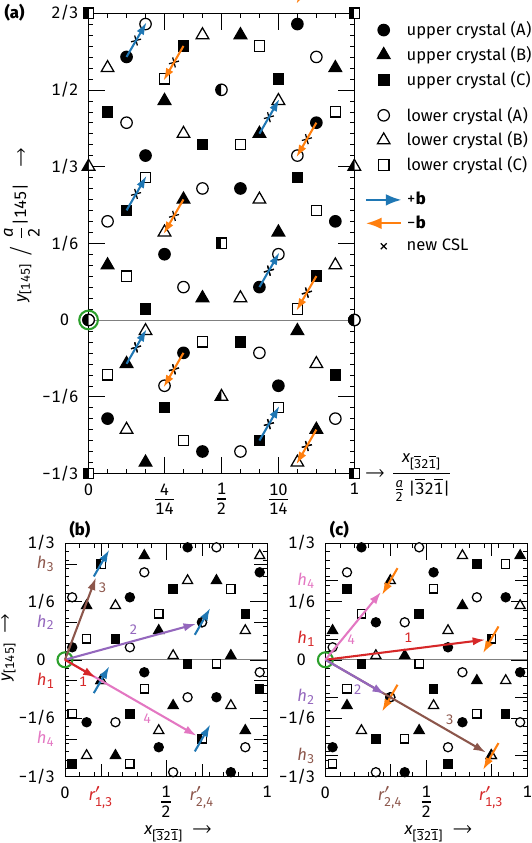}
    \caption{Obtaining disconnection properties from a dichromatic
      pattern, here exemplarily for $\Sigma$7 $[11\overline{1}]$ tilt
      GBs (a). Blue and orange arrows show two possible DSC vectors
      $\pm \mathbf{b}$. Applying $+\mathbf{b}$ or $-\mathbf{b}$
      displacements to the crystallites, yields the new patterns in
      (b) and (c), respectively. New coincidence sites lie halfway
      along the arrows indicated in (a). The arrows numbered 1--4 in
      (b) and (c) show the distance from the old to the new
      coincidence sites. Each number corresponds to a different type
      of defect, whose step height is the component of the arrow
      normal to the GB plane.}
    \label{fig:dichromatic-pattern-S7}
\end{figure}

We first analyze line defects in our GBs in terms of their Burgers
vector. The Burgers vector of a disconnection is defined as the
dislocation content of the defect without the intrinsic dislocation
content of the GB (according to, e.g., Read--Shockley
\cite{Read1950}). For this, half Burgers circuits are drawn in both
crystallites along low-index crystal directions, rotated into a common
coordinate system, and then summed \cite{Pond_1989}. The issue is that
this circuit cannot be reliably drawn in the GB core. However, the GB
structure on both sides of a disconnection is the same, so we can
choose two equivalent GB crossings across the same GB motifs that
cancel each other \cite{Medlin2017} (see also Refs.~\cite{Saba2023,
  brink2024, pemma2024} for more examples). We always draw the circuit
counter-clockwise around the $z$ axis for consistency. This allows us
to correctly evaluate the net defect content of edge dislocations that
fully span the length of the GB along the tilt axis direction.

There are some cases, where this is not possible. One example would be
the GB phase junction, which is the line defect separating two GB
phases, where the GB crossings are necessarily different on each side
of the defect \cite{Frolov2021}. Here, we can instead directly measure
lines crossing a GB segment in a defect-free reference structure
corresponding to that segment. The translation into the reference
structure uses atomic motifs as orientation points \cite{Frolov2021}.

The GB phases and their defects were analyzed and visualized with
OVITO \cite{Stukowski2009}. We did not employ the automated
Interfacial Line Defect Analysis (ILDA) method \cite{Deka2023},
because it currently requires the clear identification of an atom
within the GB as a coincidence site. Due to the complex reconstructions
of domino/zipper and pearl, it was not possible to correctly define
these sites. We therefore analyzed the defects manually as described above.

\subsection{Defect mode analysis}
\label{sec:methods:defect-mode-analysis}

For a given Burgers vector $\mathbf{b}$, the dichromatic pattern
yields additional information about the disconnection mode. To
demonstrate this, Fig.~\ref{fig:dichromatic-pattern-S7} shows a
$\Sigma7$ $[11\overline{1}]$ tilt GB with a simpler dichromatic
pattern than our $\Sigma$37c GBs. The dichromatic pattern is simply
the overlay of the atomic coordinates of the upper and lower
crystallite, such that certain atoms of both crystallites occupy the
same position (coincidence sites) \cite{Bollmann1970}. The coincidence
sites form the coincidence site lattice (CSL), which is periodic. The
possible Burgers vectors can then be drawn to connect two sites in the
dichromatic pattern, forming the displacement-shift-complete (DSC)
lattice \cite{Bollmann1970, lejcek2010grain, Han2018}.  We can
(arbitrarily) imagine the left side of the defect corresponding to the
pattern in Fig.~\ref{fig:dichromatic-pattern-S7}(a), while the right
side of the defect corresponds to a pattern that was obtained by a
relative shift $\mathbf{b}$ between the crystallites
(Fig.~\ref{fig:dichromatic-pattern-S7}(b)). To minimize overall strain
in the system, we apply a displacement of $\mathbf{b}/2$ to the upper
crystallite and $-\mathbf{b}/2$ to the lower crystallite. A new
coincidence site can thus be found in
Fig.~\ref{fig:dichromatic-pattern-S7}(a) halfway along the Burgers
vector, marked by an x. We could measure every interesting property
using this point, but also illustrate the shifted dichromatic patterns
in addition. We see that the dichromatic pattern is equivalent, but
underwent a translation. This is the (common) case when $\mathbf{b}$
is not a CSL vector.

The step height $h$ of a disconnection can be measured as the offset
normal to the GB between two crystallographically equivalent sites on
both sides of the defect. Since this offset will be the same for any
site, we choose a coincidence site for simplicity. The green circle in
Fig.~\ref{fig:dichromatic-pattern-S7} corresponds to the ``left''
coincidence site. Obviously, for a given $\mathbf{b}$, an infinite
number of possible new equivalent sites exist
(Fig.~\ref{fig:dichromatic-pattern-S7}(b)) and thus an infinite number
of possible step heights. Steps that have a component along the tilt
axis ($z$) are also possible: Defect 1 in
Fig.~\ref{fig:dichromatic-pattern-S7}(b) connects a coincidence site
on stacking plane A with one on plane B. These defects introduce an
additional $z$ step into the atomic structure of the GB, but do not
influence the GB plane, unlike the step $h$. Smaller step heights are
likely energetically favorable \cite{Pond1977b, King1980}, although
sometimes, e.g., the second smallest step can be preferred
\cite{pemma2024}. We note that inverting the Burgers vector also
inverts the step heights (Fig.~\ref{fig:dichromatic-pattern-S7}(c)),
which we will discuss in more detail in the results.

We are also interested in the minimum distance $r$ between two
disconnections. If two defects do not overlap, we must find an
equivalent site on each side of the defect that is not part of the
next defect's core. These equivalent sites are found in defect-free
structures with the periodicity $R$ of the CSL. However,
when the dichromatic pattern is shifted, an additional distance $r'$
must be considered (marked in Fig.~\ref{fig:dichromatic-pattern-S7}(b)--(c)
on the abscissa). The exact minimum distance $r$ depends on the
atomic structure of the defect, but the admissible values are $r = |n
R \pm r'|$, with $n$ being an integer.

\section{Grain boundary phases as a function of inclination}
\subsection{Structure search}

\begin{figure}
    \centering
    \includegraphics[width=\linewidth]{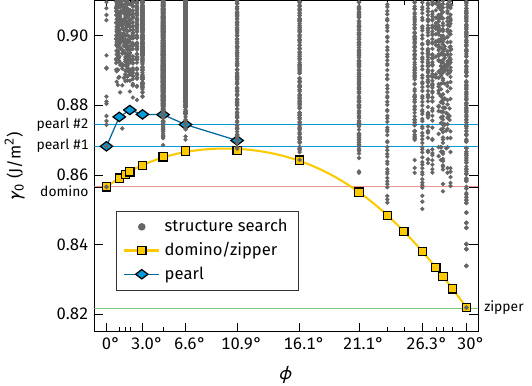}
    \caption{Ground state energies $\gamma_0$ of GBs found via
      structure search, construction of low-energy domino/zipper GBs,
      or by annealing simulations. The data is presented as a function
      of the GB inclination $\phi$ from the symmetric reference plane
      (1~10~11). The second symmetric plane, (437), is located at
      $\phi = \ang{30}$. The energies of the symmetric GB phases are
      indicated with horizontal lines.}
    \label{fig:ground-state-energy}
\end{figure}

We used different strategies to search for possible GB phases with
inclinations from \ang{0} to \ang{30}. First, we constructed
domino/zipper GBs by starting from the domino structure and
continually increasing one facet \cite{brink2024}. This yields
periodic simulation cells with a range of finite inclinations. Using
the crystallographic planes obtained with this method, we also started
a structure search with \textsc{grip} for the same inclinations. These
calculations each yield dozens to hundreds of GB structures, many of
them defective. Figure~\ref{fig:ground-state-energy} shows the
resulting GB energies $\gamma_0$ of the ground state. For small
inclinations the periodic cell length is comparatively large and
\textsc{grip} cannot easily find low-energy structures. For the
symmetric GBs and $\phi = \ang{6.59}, \ang{10.89}$, however,
\textsc{grip} yields good results. It still finds mostly reasonable
structures in the range $\ang{4.72} \leq \phi \leq \ang{23.41}$. We
find that the ground state GB energy of the domino/zipper phase is
always the lowest one, although it cannot always be found with
\textsc{grip}.

Additionally, we extracted the lowest-energy pure pearl
structures. Non-faceted pearl phases only exist up to
\ang{10.89}. Beyond that, we find either the domino/zipper phase or a
mix of zipper and pearl facets. For \ang{1} to \ang{3}, the
\textsc{grip} structures have relatively high GB energies, and we
found lower-energy pearl structures by following an annealing
procedure. In that case, we annealed a starting domino/zipper
structure for \SI{100}{ns} at \SI{400}{K} and then minimized the
result. Consequently, the pearl data points in
Fig.~\ref{fig:ground-state-energy} for low $\phi$ come from the
annealing simulations.

It stands to reason that the constructed domino/zipper GBs are the
overall lowest-energy structures. Neither structure search with
\textsc{grip}, nor annealing yielded lower-energy structures in any
case.
For the pearl phase, the structures we found have GB energies very
close to the symmetric pearl phase. We will discuss the structures and
their defects in detail in Sec.~\ref{sec:analyze-defects} and first
focus on the stability of these GB phases as a function of
temperature.

\subsection{Excess free energy}

\begin{figure}
    \centering
    \includegraphics[width=\linewidth]{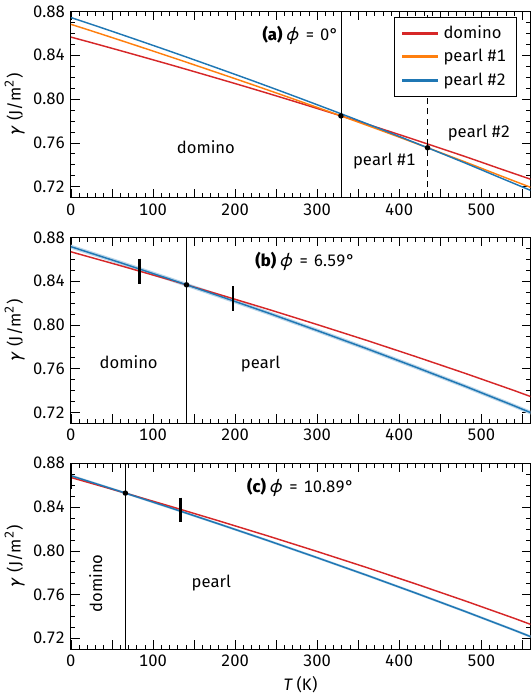}
    \caption{Excess free energy calculations for pearl and domino GB
      phases at some representative inclinations $\phi$. The predicted
      GB phase transitions and stability ranges are marked. For the
      asymmetric pearl GBs (blue lines in (b),(c)), different ways to
      subtract the surface free energy (see
      Eqs.~\ref{eq:gamma-asymm-1} and \ref{eq:gamma-asymm-2}) yield
      slightly different results. This is likely due to the limited
      simulation cell size, where strain fields of GB and surface
      slightly overlap. The estimated range of possible values is
      indicated by the blue shaded areas. The vertical marks represent
      the resulting uncertainty of the transition temperature.}
    \label{fig:qha}
\end{figure}

Free energy calculations with the QHA are only possible for smaller
simulation cells, since the eigenfrequency calculations require large
amounts of memory and CPU time. Thus, we compared domino and pearl
phases for $\phi = \ang{0}$, \ang{6.59}, and \ang{10.89}, where
low-energy structures with small periodic lengths along $y$ could be
obtained with \textsc{grip}. Figure~\ref{fig:qha} shows the
results.\footnote{Some of us reported a higher transition temperature
  of \SI{460}{K} for the symmetric case in an earlier work
  \cite{Langenohl2022}. In a later work \cite{Brink_2023} and in the
  present case, we used a bigger cell for the free energy calculations
  and obtain \SI{330}{K} instead. This highlights the sensitivity of
  these results to small numerical errors due to the very similar
  slope of the free energy curves. We will discuss the match to
  annealing simulations and experiment further down.}  Similar to the
GB energies at \SI{0}{K} (Fig.~\ref{fig:ground-state-energy}), the
excess free energy of the pearl phases is only slightly higher for the
asymmetric GBs than for the symmetric GB. The domino phase, however,
exhibits a steadily increasing excess free energy with higher
inclinations $\phi$. As a consequence, the simulations predict that
the domino phase is only stable at very low temperatures in asymmetric
GBs. A reasonable hypothesis is that domino and pearl exhibit
different kind of defects with different energies in asymmetric
GBs. We will analyze these defects in detail later in
Sec.~\ref{sec:analyze-defects}.

\subsection{Annealing simulations}

\begin{figure}[t!]
    \centering
    \includegraphics[width=\linewidth]{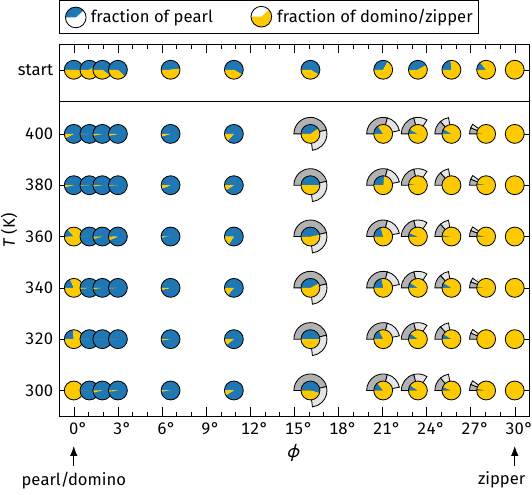}
    \caption{The GB phases present in asymmetric tilt GBs after
      equilibrating at temperatures between $\SI{300}{K}$ and
      $\SI{400}{K}$ for $t = \SI{100}{ns}$. The topmost row shows the
      fraction of GB phases in the starting structure, which were
      structures obtained with GRIP that contain both pearl and
      domino/zipper. For $\phi = \ang{0}$, we constructed a box with
      one big region each for pearl and domino. Results with
      additional starting structures are provided in Supplemental
      Fig.~\ref*{fig:suppl:phase-fractions}. For each inclination and
      temperature, the pie chart shows the fraction of the two GB
      phases. Starting from \ang{16}, there are gray bars around the
      data points. These indicate the maximum possible fractions of
      pearl phase in faceted GBs (see
      Sec.~\ref{sec:analyze-defects:high-incl}). The dark gray region
      is for symmetric pearl facets. The light gray region indicates
      how much longer the pearl facets can become if they are
      asymmetric ($\phi = \ang{10.89}$).}
    \label{fig:phase-fractions}
\end{figure}

In order to check the QHA results over a larger range of inclinations,
we performed several annealing simulations. First, we selected
defective \textsc{grip} structures that contained both domino and
pearl. For each inclination, we chose two different cells and annealed
each structure for \SI{100}{ns} at a range of temperatures
(Fig.~\ref{fig:phase-fractions} and Supplemental
Fig.~\ref*{fig:suppl:phase-fractions}(a)). We also started from the pure
domino/zipper phase (Supplemental
Fig.~\ref*{fig:suppl:phase-fractions}(b)). Finally, we used the outcome
of these simulations at \SI{400}{K} that contained a large amount of
pearl, for which we restarted the annealing process at different
temperatures (Supplemental Fig.~\ref*{fig:suppl:phase-fractions}(c)).

The results for $\phi \leq \ang{10.89}$ mostly follow the predictions
from the QHA calculations. Domino is only stable in this temperature
range for $\phi = \ang{0}$. In these symmetric GBs, domino transforms
into pearl above 340 to \SI{360}{K}, close to our prediction using the
QHA. Below that temperature range, the results depend on the starting
condition due to the slow kinetics and mixed structures can
occur. Nevertheless, pure pearl GBs start transforming into domino
(Supplemental Fig.~\ref*{fig:suppl:phase-fractions}(c)), which
indicates that the QHA calculations (Fig.~\ref{fig:qha}) correctly
predict the transition temperature. The fact that we often obtain
metastable mixed structures highlights the small free energy
differences of the GB phases in this temperature range.
When starting from a pure domino structure in asymmetric GBs, the
domino phase remains in the system at low temperatures (Supplemental
Fig.~\ref*{fig:suppl:phase-fractions}(b)). This evidences the
relatively slow kinetics of GB phase transitions and the resulting
hysteresis of the transition. The simulations starting from mixed
pearl/domino phase are more revealing: the domino phase disappears for
$\ang{0} < \phi \leq \ang{10.89}$ (Fig.~\ref{fig:phase-fractions}). We
conclude that simulations initially containing a mix of domino/zipper
and pearl phases very likely contain equilibrated GBs, because we can
clearly observe changes of the GB phase fractions and do not require
nucleation. This is aided by the thin size (\SI{6.3}{\angstrom}) in
the periodic $z$ direction, which minimizes the barriers for the
movement of line defects. We also tried temperatures below
\SI{300}{K}, but we no longer observed any significant structural
changes within the simulation timeframe and therefore excluded these
simulations.

Above \ang{10.89}, GBs contain both pearl facets, as well as
domino/zipper facets. This indicates that the stabilization of pearl
is no longer favorable compared to faceting. To understand this
behavior, we need to analyze the defects responsible for the
inclination.

\section{Analysis of the grain boundary defects}
\label{sec:analyze-defects}

\subsection{Low inclination angles}
\label{sec:defects:low-phi}

\begin{figure}[t!]
    \centering
    \includegraphics[width=\linewidth]{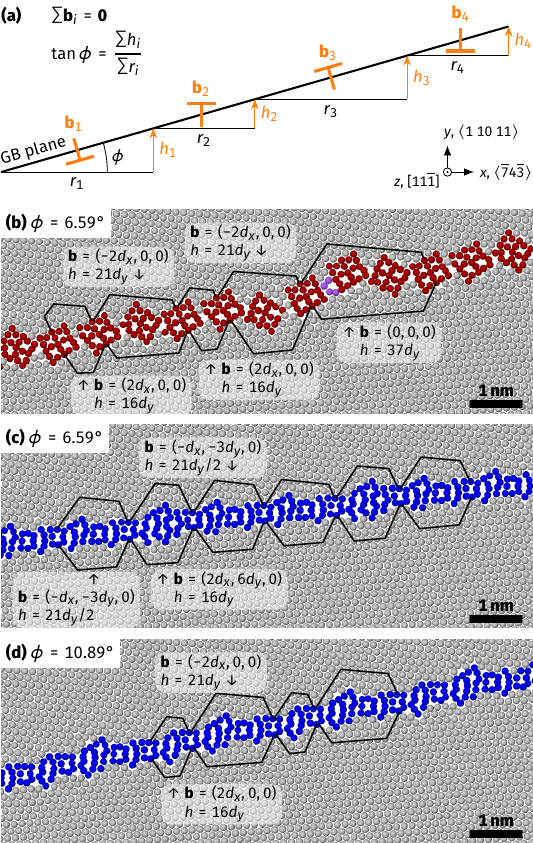}
    \caption{Compensation of the inclination at the atomic scale by
      disconnections and steps. (a) Schematic of different
      disconnection modes (consisting of Burgers vector \textbf{b} and
      step height $h$) with different slip systems present in
      asymmetric GBs. (b) Domino structure from an annealing
      simulation at \SI{300}{K}. It contains disconnctions in addition
      to pure steps. (c)--(d) Low-energy pearl structures from
      GRIP. The indicated disconnections repeat periodically to
      achieve the inclination. Note that in all cases the Burgers
      vectors add up to zero on short length scales to minimize energy
      and to preserve the misorientation.}
    \label{fig:inclination-small}
\end{figure}

In order to understand the change of stability, we systematically
looked at the GB structures we obtained and analyzed the type of
defects present in the GBs. Contrary to symmetric GBs, asymmetric GBs
are composed of pre-existing defects to compensate the deviation of
the inclination angle. Figure~\ref{fig:inclination-small} shows an
illustration of the possible defects that lead to asymmetric
inclinations without faceting. Either pure steps or disconnections
(consisting of a Burgers vector plus a step) are found. The total
inclination is a result of the sum of step heights $h$ and the
distance $r$ between defects:
\begin{equation}
  \tan \phi = \frac{\sum_i h_i}{\sum_ir_i}.
\end{equation}
In order to preserve the misorientation between the crystallites, we
require that
\begin{equation}
  \sum_i \mathbf{b}_i = \mathbf{0},
\end{equation}
summing over all defects with Burgers vectors \textbf{b} in the
GB. Otherwise, the misorientation between the abutting grains would
change \cite{Wang_2014, Tian2024GR} (cf.\ the Frank--Bilby equation
\cite{frank1950resultant, bilby1955types}). We note already from
Fig.~\ref{fig:inclination-small} that the total Burgers vector seems
to be compensated by alternating different disconnection modes
$(\mathbf{b},h)$.

\begin{figure*}[t!]
    \centering
    \includegraphics[width=\linewidth]{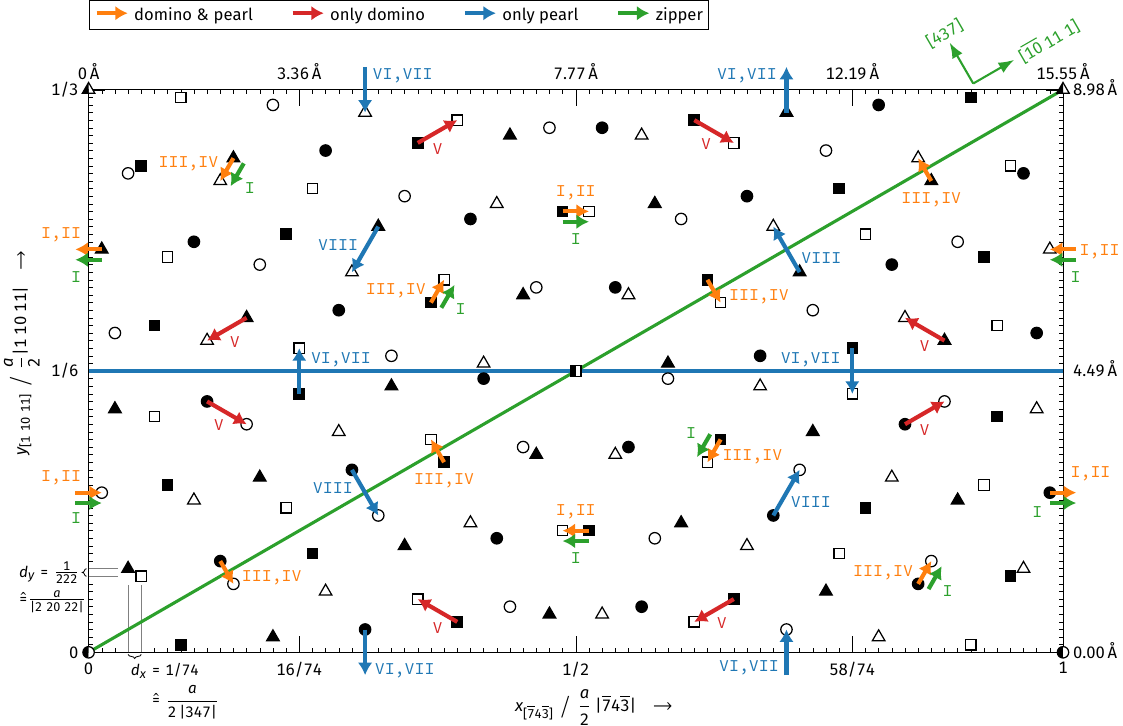}
    \caption{Part of the dichromatic pattern of $\Sigma$37c
      $[11\overline{1}]$ tilt GBs. The full pattern is shown in
      Supplemental Fig.~\ref*{fig:suppl:dichrom-S37}. Arrows indicate
      the Burgers vectors of the observed disconnections. The
      horizontal blue line schematically indicates the GB plane of the
      pearl and domino GB phases, the inclined green line indicates
      the plane of the zipper GB phase. The ticks on both axes
      represent the lattice spacings $d_x$ and $d_y$. The fractions on
      the left and bottom axes refer to the periodic lengths of the
      CSL lattice.}
    \label{fig:complexionsdefects}
\end{figure*}

We investigated our asymmetric tilt GBs at different temperatures and
inclination angles and compiled a list of the disconnection types
observed in each GB phase. Figure~\ref{fig:complexionsdefects} shows
the observed Burgers vectors in the dichromatic pattern. We also
measured the dislocation spacing $r$ along the GB plane by choosing
the two closest equivalent sites left and right of the disconnection
core and measuring their projected distance along the symmetric GB's
$x$ direction. We compared this with the admissible spacings from the
dichromatic pattern (see methods in
Sec.~\ref{sec:methods:defect-mode-analysis}).

We categorize the disconnection types by their mode $(\mathbf{b}, h)$,
see Fig.~\ref{fig:complexionsdefects}. Additionally, we differentiate
between disconnections with a zero $b_{y}$ component normal to the GB
plane (glide disconnections) and disconnections with nonzero $b_{y}$
component (climb disconnections). This is because the slip plane of
the glide disconnection is equal to the GB plane and they are
consequently mobile. The climb disconnections possess a slip plane
that is not equivalent to the GB plane. Since disconnections cannot
exist outside the GB, these disconnections require disconnection
climb---and thus diffusion---to move along the GB \cite{Ashby1969,
  balluffi2005kinetics, Han2018}.  Glide disconnections are generally
found in symmetric GBs during shear-coupled GB migration
\cite{Cahn2006, Rajabzadeh2013, Han2018, Larranaga2020}.

\begin{table}[b!]
\centering
\caption{Comparison of inclinations obtained by representative
  combinations of disconnections that lead to an overall Burgers
  vector of zero, but a nonzero overall step height $h$. Type S is a pure
  step; see Fig.~\ref{fig:complexionsdefects} for the other individual
  disconnection types. There is a minimum possible spacing
  $r_\text{combined}$ between one combined defect and the next,
  similar one. It is used to calculate the maximum inclination
  $\phi_\text{max} = \arctan(h/r_\text{combined})$. Note that
  $d_y/d_x = 1/\sqrt{3}$.}
\label{tab:GBdefectsheight}
\begin{tabularx}{\linewidth}{Xcccc}
  \toprule
  \multicolumn{2}{l}{Defect combination}
  & $h$ & $r_\text{combined}$ &  $\phi_\text{max}$ \\
  \midrule
  Domino &  S       &  $n\cdot37d_y$ & $(1+n/2)74d_x$  &  0°--30°     \\
         & I+II     &  $37d_y$ & $111d_x$       & 10.89°        \\
         & III+IV   &  $37d_y$ & $111d_x$       & 10.89°        \\
         & I+IV+V   &  $37d_y$ & $132d_x$       &  9.19°        \\
         & II+III+V &  $37d_y$ & $185d_x$       &  6.59°        \\
  \midrule
  Pearl  & I+II     &  $37d_y$ & $111d_x$       & 10.89°        \\
         & III+IV   &  $37d_y$ & $111d_x$       & 10.89°        \\
         & VI+VII   &  $37d_y$ & $111d_x$       & 10.89°        \\
         & 2$\times$VII & $74d_y$ & $116d_x$    & 20.22°        \\
         & 2$\times$III+VIII & $37d_y$ & $185d_x$ &  6.59°        \\
  \midrule
  Zipper &  S     & $n\cdot37d'_y$ & $(1+n/2)74d'_x$   &  0°--30°     \\
  \bottomrule
\end{tabularx}
\end{table}

\begin{figure}[t!]
    \centering
    \includegraphics[width=\linewidth]{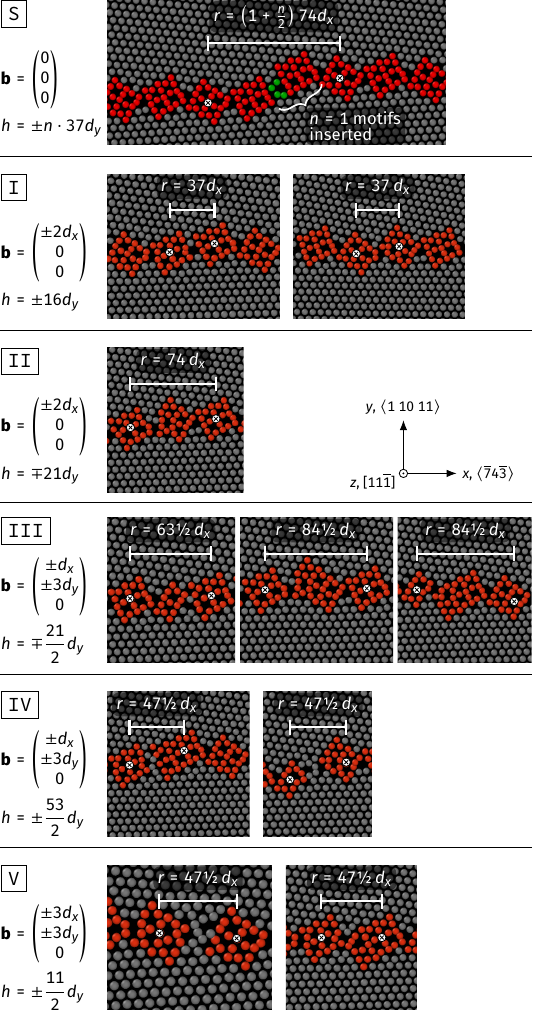}
    \caption{The disconnection types observed in asymmetric domino
      GBs. The sign of $h$ depends on the sign of $b_x$ (either the
      same or the opposite sign), while the sign of $b_y$ is
      independent. The disconnection spacing $r$ was measured as the
      projected distance along $x$ between the two closest equivalent
      sites adjacent to the disconnection core (marked atoms).}
    \label{fig:defects-domino}
\end{figure}
\begin{figure}[t!]
    \centering
    \includegraphics[width=\linewidth]{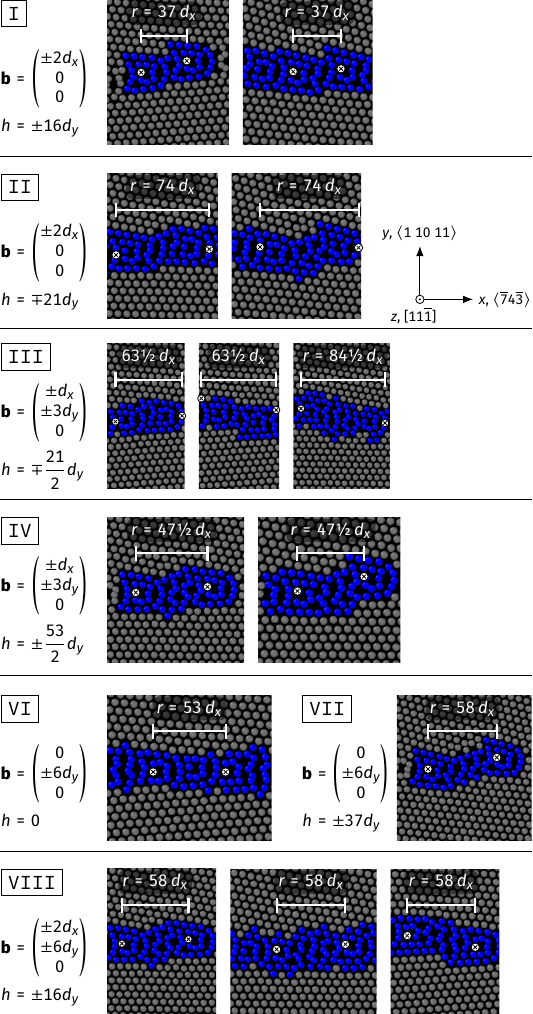}
    \caption{The GB disconnection types observed in pearl of
      $\Sigma$37c asymmetric grain boundaries. The signs of $h$ depend
      on the sign of $b_x$ (either same or opposite), while the sign
      of $b_y$ is independent. This means that for type VII, both
      signs for the step are always possible.}
    \label{fig:defects-pearl}
\end{figure}

The GB disconnections observed in domino and pearl are shown in
Figs.~\ref{fig:defects-domino} and
\ref{fig:defects-pearl}. We express the Burgers vectors and step
heights in units of the lattice spacings (see also
Fig.~\ref{fig:complexionsdefects}), which are
\begin{align}
  d_x &= \frac{a}{2 \cdot |347|} \approx \SI{0.21}{\angstrom} \\
  d_y &= \frac{a}{2 \cdot |1~10~11|} \approx \SI{0.12}{\angstrom} \\
  d_z &= \frac{a}{|111|} \approx \SI{2.09}{\angstrom},
\end{align}
where $|hkl|$ is the length ($L_2$-norm) of the vector
$(h,k,l)$. Generally, the disconnection types I, II, III, and IV are
the most commonly observed ones. They have the shortest Burgers vector
(\SI{0.42}{\angstrom}) and thus also a low defect energy. Types I/II
and III/IV differ by their orientation towards the GB plane, with I
and II being glide disconnections and III/IV being climb
disconnections. As discussed earlier, domino can also have pure steps,
here named type S.

Our analysis shows that the sign of the step height $h$ is dependent
on the Burgers vector component $b_x$ (see Supplemental,
Sec.~\ref*{sec:suppl:cryst-poss-def}). This means that when a single
disconnection type $(\mathbf{b}_1, h_1)$ is used, no total step height
can be achieved while conserving overall $\mathbf{b} = \mathbf{0}$,
because $-\mathbf{b}_1$ leads to $-h_1$. Different modes, in contrast,
can be summed up such that the overall Burgers vector is zero, but the
overall step height is not:
$(\mathbf{b}_1, h_1) + (\mathbf{b}_2, h_2) = (\mathbf{0}, h)$. We find
that I and II can have the same sign of the step height when their
Burgers vectors have opposite signs. Thus, they can compensate each
others Burgers vector while leaving behind a step. These pairs are for
example observed in Fig.~\ref{fig:inclination-small} for domino and
for pearl at $\phi=\ang{10.89}$. Type III and IV work the same
way. At higher inclination angles and temperatures, defects with
larger Burgers vectors are also observed. We found type V only in
domino and types VI--VIII only in pearl. Interestingly, the
combination of two times type III with one type VIII defect leads to a
low-energy pearl structure at $\phi = \ang{6.59}$
(Fig.~\ref{fig:inclination-small} shows the structure obtained with
\textsc{grip}).

\begin{figure}[t!]
    \centering
    \includegraphics[width=\linewidth]{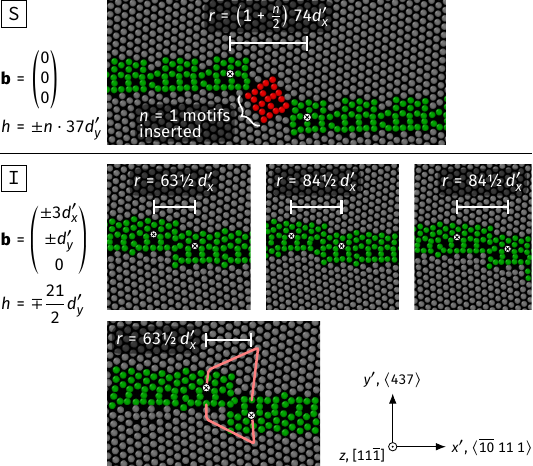}
    \caption{The step and disconnection observed in asymmetric zipper
      GBs. Values are expressed in the indicated coordinate system,
      which is aligned with the GB plane and therefore differs from
      Figs.~\ref{fig:defects-domino}--\ref{fig:defects-pearl}. The
      sign of $h$ depends on the sign of $b_x$ (either same or
      opposite), while the sign of $b_y$ is independent. The step S is
      equivalent to the steps in domino, only now rotated onto the
      zipper GB plane. In the last snapshot, the zipper motifs on the
      left and right of the defect are mirrored, so that the
      equivalent sites are once on top and once on the bottom of the
      square motif. This is expected, there are two degenerate states
      of the zipper structure that are sheared either to the left or
      right \cite{brink2024}. This additional shear must cancel out in
      the Burgers circuit (red lines). We verified that this is true
      also with the Frolov et al.\ method \cite{Frolov2021} for the
      Burgers circuit.}
    \label{fig:zipper_defectsATGB}
\end{figure}

A list of several defect combinations is provided in
Table~\ref{tab:GBdefectsheight}. It is not exhaustive, but contains
representative low-energy configurations. Given the observed minimum
distances $r$ between individual defects
(Figs.~\ref{fig:defects-domino}--\ref{fig:zipper_defectsATGB}), we
calculate the minimum distance $r_\text{combined}$ between defect
combinations as the sum of $r$. We can thereby estimate the maximum
inclination change that can be achieved by these disconnection
combinations as $\phi_\text{max} = \arctan(h/r_\text{combined})$. If
the distance between defects is larger than $r_\text{combined}$, the
inclination change is smaller. We see that typically the range of
inclinations is limited to at most \ang{11}, except for pure steps in
domino. This fits to the earlier observations, where pure pearl
structures were not found above $\phi = \ang{10.89}$, neither when
annealing nor with the structure search method. Instead, faceted
structures were found. There is one defect combination of two times
type VII for pearl that could hypothetically achieve
$\phi = \ang{20.22}$, but we did not observe this in our simulations,
most likely due to the high defect energy.

To connect this to the relative stability, we can estimate the line
energy of the defect combinations from $\gamma_0(\phi)$
(Fig.~\ref{fig:ground-state-energy}). For this, we project the GB
energy onto the symmetric GB plane and normalize by the defect
distance $r$:
\begin{equation}
  \frac{E_\text{def}}{t} = r
  \left(\frac{\gamma_0(\phi)}{\cos\phi} - \gamma_0(\phi{=}0)\right),
\end{equation}
where $E_\text{def}$ is the defect energy and $t$ is the line length of
the defect combination (i.e., the sample thickness). Using the values
in Table~\ref{tab:GBdefectsheight}, we obtain
\begin{align*}
  \text{domino: }&                 \SI{62}{pJ/m} \text{ (S, $n = 1$)} \\
  \text{pearl \phantom{0}6.59°: }& \SI{47}{pJ/m} \text{ (2$\times$III+VIII)} \\
  \text{pearl 10.89°: }&           \SI{41}{pJ/m} \text{ (I+II)}
\end{align*}
(see Supplemental Fig.~\ref*{fig:suppl:combined-defect-energies} for
more details). The values for domino depend only very weakly on the
inclination (changes $< \SI{2}{pJ/m}$ from \ang{1} to \ang{11}). For
the same resulting step height ($37d_y$), the combined pearl defects
cost less energy than the steps in domino, thereby stabilizing the
asymmetric pearl phases.

In some cases, phase junctions \cite{Frolov2021} between domino and
pearl occur (Supplemental Fig.~\ref*{fig:suppl:phase-junc}). They also
have a Burgers vector \cite{Frolov2021}. For a pure phase junction
without any adjacent disconnections, we find a Burgers vector of
approximately
$(\SI{0.20}{\angstrom}, \SI{0.08}{\angstrom}, \SI{0.62}{\angstrom})$
in accordance with the values obtained in previous observations
\cite{Langenohl2022}, see Supplemental
Fig.~\ref*{fig:suppl:phase-junc}. For $\phi \leq \ang{11}$, pearl and
domino should not coexist in thermodynamic equilibrium. While GB phase
patterning can exist, it is not possible in perfect $\Sigma37$c tilt
GBs due to their high junction energies \cite{Winter2024}. Experimentally
observed patterning in this GB type is either due to additional
defects to compensate a twist or tilt component, or due to the slow GB
phase transition kinetics \cite{Meiners2020, Langenohl2022}. Here, the
phase junctions are thus not relevant for the stability of the GB
phases in pure tilt boundaries. They are artifacts of the slow GB
phase transition kinetics close to room temperature.

We also studied the disconnections in the zipper GB phase, which lies
on the other symmetric GB plane
(Fig.~\ref{fig:zipper_defectsATGB}). We examined these defects in the
coordinate system aligned with the GB plane and therefore express
values in terms of the lattice spacings $d'_x = d_y$ and $d'_y =
d_x$. Zipper structures contain fewer defects than the pearl and
domino structures and we only found disconnections of type I, which is
a climb defect. Glide defects in zipper would have much larger Burgers
vectors (cf.\ Fig.~\ref{fig:complexionsdefects}) and do not seem to
occur here. We note that combinations of the observed disconnection cannot
both lead to $\mathbf{b} = \mathbf{0}$ and $h \neq 0$, so inclination
changes are solely due to pure steps. These steps are the same as for
domino.

\subsection{Experimental evidence}

\begin{figure}[t!]
    \centering
    \includegraphics[width=\linewidth]{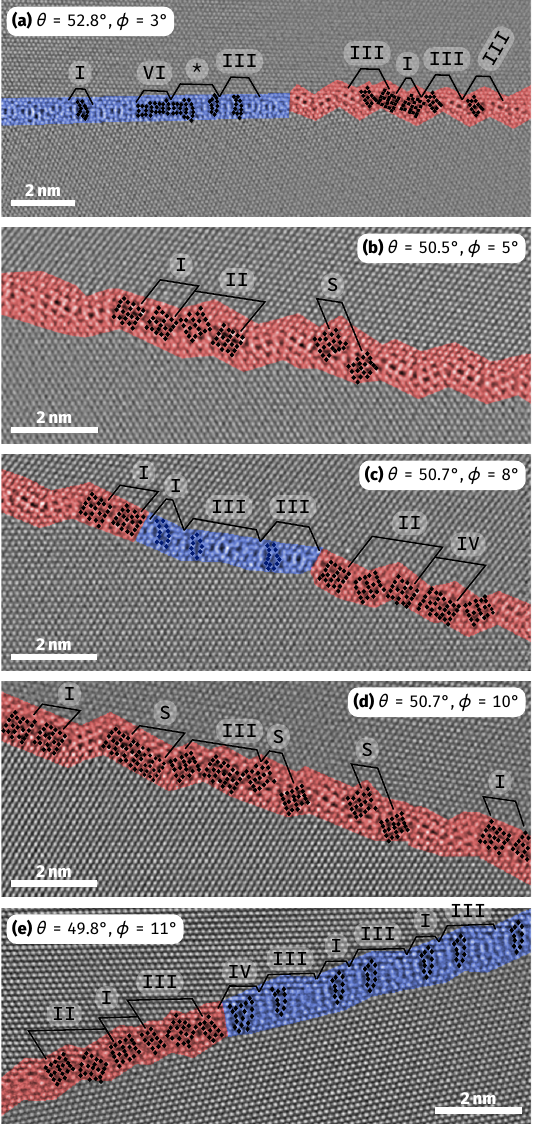}
    \caption{HAADF-STEM images of asymmetric GBs in Cu. Motifs and
      disconnection types are marked. GB phases are colored to
      indicate domino (red) and pearl (blue). Raw images are provided
      in Supplemental Fig.~\ref*{fig:suppl:Cu_exptimages}. The
      asterisk in (a) marks a segment with no overall defect character
      and looks like a dipole of two type I defects with opposite
      Burgers vector and step height.  Note that (a) and (e) deviate
      from the perfect misorientation angle $\theta = \ang{50.6}$ and
      the defects add up to an overall nonzero $b_y$ component to
      compensate this. The misorientation does not deviate in (d),
      although the visible Burgers vectors do not cancel out. However,
      there are many steps that do not exist in (a) and (e), so it is
      likely that the defects are outliers that are compensated
      outside the visible segment.}
    \label{fig:Cu_exptimages}
\end{figure}

HAADF-STEM images of several asymmetric GBs in Cu are shown in
Fig.~\ref{fig:Cu_exptimages}. We first note that a significant amount
of domino phase is still found in these GBs, while the simulations
predict that asymmetric domino phases at inclinations
$\phi < \ang{11}$ are only stable far below room temperature. This
could indicate that the potential overestimates the stability of the
pearl phase. However, the experimental GBs also differ from the
idealized, simulated GBs in several ways. If the GB deviates from the
$\Sigma37$c misorientation or contains small twist components, the GB
phase stability will be affected and even domino/pearl phase
patterning could occur \cite{Langenohl2022}. Stresses and strains in
the thin film or due to nearby triple junctions will also change the
stability ranges of the GB phases \cite{Meiners2020}, especially close
to the transition point, where the excess free energy differences are
small. These stresses are temperature dependent, so the occurrence of
domino phase in the experimental samples is a nontrivial function of
temperature. Finally, metastable domino phase may occur due to thermal
fluctuations during deposition or annealing. The experimental samples
are much thicker than the simulated ones, so we expect the GB
transformation kinetics to be much slower \cite{Meiners2020}. We
therefore cannot exclude that GB phases in experimental samples are
simply prevented from transforming kinetically.

We nevertheless observe steps in domino and disconnection types I--IV
in both GB phases, which confirms the simulation result that these are
the most common defects with the smallest Burgers vectors. We also
find the type VI defect in pearl, but not in domino. The types I/II and
III/IV often appear in pairs (as discussed above), sometimes paired
across different GB phases. In particular in
Fig.~\ref{fig:Cu_exptimages}(a) and (e), pairs of types I and III are
observed without many compensating defect types II or IV. While the
STEM images only cover a small part of the overall GB, the observed
combinations add up to an overall nonzero Burgers vector, at least
locally. We measured the misorientation of these GBs and found small
deviations from the ideal value $\theta = \ang{50.6}$. The net Burgers
vector component $b_y$ is thus responsible for this (local) deviation.

\begin{figure}[t!]
    \centering
    \includegraphics[width=\linewidth]{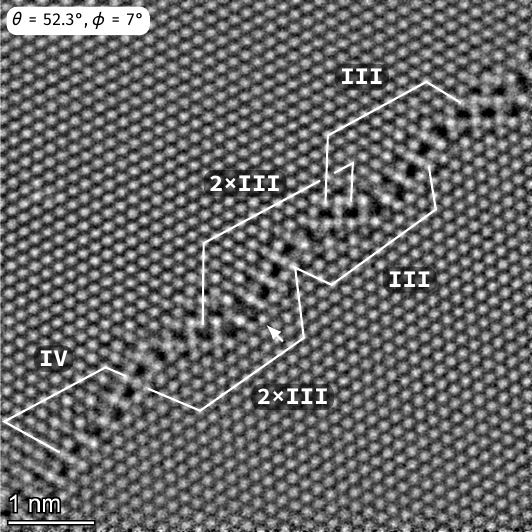}
    \caption{HAADF-STEM image of an asymmetric GB in
      Al. Disconnections are marked. The misorientation deviates from
      $\theta = \ang{50.6}$ for $\Sigma$37c, so instead of pure steps
      type III and IV defects are combined such that a $b_y$ component
      remains (see text). More images with similar defects are shown
      in Supplemental Fig.~\ref*{fig:suppl:Al_exptimages}.}
    \label{fig:Al_exptimages}
\end{figure}

We also present data from an Al thin film in
Fig.~\ref{fig:Al_exptimages}. For this material, we know that the
pearl phase is not observed \cite{Brink_2023, Saba2024}, so the data
is limited to the domino/zipper phase. The misorientation also
deviates more from $\Sigma 37$c than in the Cu samples and our data
only includes shorter segments.

We can nevertheless identify the defect types III and IV in these GBs,
in accordance with the expectations from the simulation. This defect
combination can lead to a net zero Burgers vector overall, as
discussed before, but since $b_y$ can have either sign, they can also
add up to $b_x = 0$ but $b_y \neq 0$. The latter is the case here and
it is likely the reason that disconnections instead of steps are
observed: the deviation from the perfect $\Sigma 37$c misorientation
is compensated by this $b_y$ component. The type III defects manifest
here as segments with three square motifs instead of two. We also
observed a $2\times$III defect (double the Burgers vector and step
height of type III), with three clearly visible square motifs and a
less clearly-defined motif at the junction (arrow).

Overall, the predicted structures and defect types in
Sec.~\ref{sec:defects:low-phi} agree well with the present
experimental observations. However, the low predicted transition
temperatures in Cu (Figs.~\ref{fig:qha} and \ref{fig:phase-fractions})
cannot explain the experimental observation of the domino phase at
room temperature. This might be because the interatomic potential
predicts the wrong free energies, but importantly the transformation
kinetics and the actual stress state also strongly influence GB phase
occurrence in real samples.

\subsection{High inclination angles and faceting}
\label{sec:analyze-defects:high-incl}

\begin{figure}[t!]
    \centering
    \includegraphics[width=\linewidth]{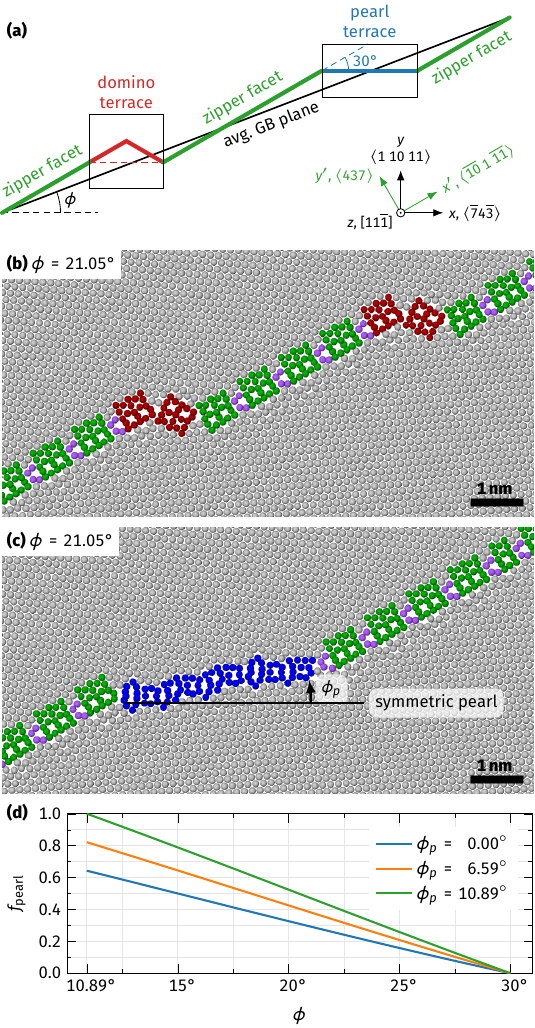}
    \caption{Inclination angles with $\ang{11} < \phi < \ang{30}$ lead
      to faceted GBs. (a) When referencing to $\phi = \ang{0}$, we can
      also describe one facet as a terrace defect. There are GB phase
      junctions between the facet and the terrace. (b) Due to elastic
      interactions, the domino facets have minimal length (see
      text). (c) Pearl facets have a driving force to grow to minimize
      the number of GB phase junctions/facet junctions. The pearl
      facets, however, also exhibit defects to increase their
      inclination $\phi_p$ from the symmetric $\{1~10~11\}$
      plane. This is due to the low energy of the pearl defects
      (cf.~low $\phi$ in Fig.~\ref{fig:ground-state-energy}). (d) For
      a mix of pearl and zipper, we can geometrically estimate the
      fraction of pearl facets. This fraction will increase if the
      pearl phase is additionally inclined by $\phi_p$ due to its own
      defects.}
    \label{fig:terrace-facet}
\end{figure}

Beyond $\phi \approx \ang{11}$, the pearl phase can no longer
compensate the inclination by introducing disconnections, while the
domino phase can still have steps. The GBs with higher inclination
angles must therefore either be pure domino/zipper phases or they must
facet. In fact, we can also think of steps in domino/zipper as
faceting (Fig.~\ref{fig:terrace-facet}(a)). We notice, however, that
designating steps in zipper as domino phase facets leads to very small
domino facets that rather resemble steps
(Fig.~\ref{fig:terrace-facet}(b)). Instead of extended domino facets,
individual step-like defects are preferred. This is because this
configuration minimizes the strain energy (see Supplemental
Fig.~\ref*{fig:suppl:domino-terrace-stress}) and domino/zipper facet
junctions do not cost energy \cite{brink2024}. For pearl facets
(Fig.~\ref{fig:terrace-facet}(c)), the segments are extended, because
zipper/pearl phase junctions are connected with an energy cost. The
pearl facet is often not defect free and deviates from
$\phi = \ang{0}$.

Geometrically, the amount of pearl facets can be calculated as the
fraction $f_\text{pearl}$ of pearl along the average GB plane (see
Supplemental, Sec.~\ref*{sec:suppl:facet-fractions} for details). This
depends on the inclinations of the two facets via
\begin{equation}
  \label{eq:pearl-facet-frac}
  f_\text{pearl} = \frac{a_z\cos\phi - \cos\phi_z}
                        {a_z\cos\phi_p - a_p\cos\phi_z} a_p,
\end{equation}
with
\begin{align}
  a_p &= \cos\phi_p\cos\phi + \sin\phi_p\sin\phi \\
  a_z &= \cos\phi_z\cos\phi + \sin\phi_z\sin\phi.
\end{align}
Here, $\phi$ is the average inclination of the GB plane, $\phi_p$ is
the inclination of the pearl facets, and $\phi_z$ is the inclination
of the zipper facets. As discussed in Sec.~\ref{sec:defects:low-phi},
zipper is almost defect-free and steps in zipper can be thought of as
domino phase. We thus assume a fixed $\phi_z = \ang{30}$. For pearl,
inclinations up to \ang{10.89} are possible. We thus plotted three
different values of $\phi_p$ in Fig.~\ref{fig:terrace-facet}(d). For
$\phi = \ang{16.10}$, the pearl fraction indeed corresponds to the
predicted value (Fig.~\ref{fig:phase-fractions}, gray bars). For
inclinations at \ang{21.05}, pearl still exists, but the fraction of
pearl phase is lower than predicted. Instead, domino/zipper facets
appear partially in their place. At even higher inclinations, we find
little to no pearl phase. The GB energies in
Fig.~\ref{fig:ground-state-energy} explain this. The energy of the
domino/zipper phase decreases sharply for higher inclinations, but the
pearl facet energy should still correspond to the case of $\phi =
\ang{10.89}$. The increasing energy difference thus destabilizes
the pearl phase, while the region around $\phi = \ang{21.05}$
represents a transition zone where metastable mixed structures are
observed.

\begin{figure*}
    \centering
    \includegraphics[width=\linewidth]{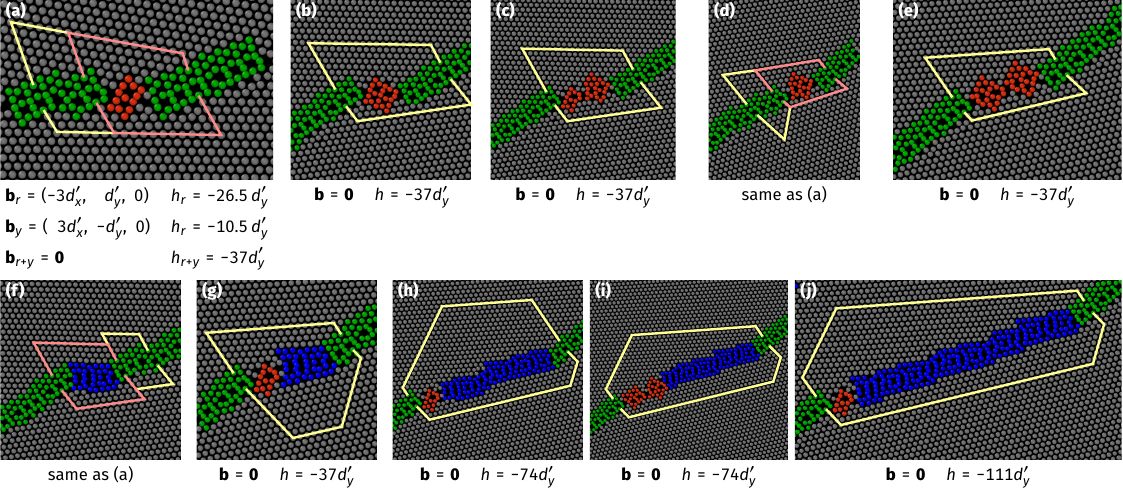}
    \caption{Analysis of some terrace--facet defects. In the panels
      with only a yellow Burgers circuit, the terrace overall acts as
      a step defect. Sometimes, there is a defect content in the
      terrace (red circuits, $\mathbf{b}_r$, $h_r$). This is then
      directly compensated by a nearby zipper type I disconnection
      (yellow circuit next to red circuit, $\mathbf{b}_y$, $h_y$). The
      terraces' disconnection mode ($\mathbf{b}_y$, $h_y$) was not
      observed in pure zipper (Fig.~\ref{fig:zipper_defectsATGB}).
      Domino terraces have minimal size to minimize their elastic
      energy (see text).}
    \label{fig:terrace-facet-defects}
\end{figure*}

In order to characterize the faceted structures in terms of GB
defects, we can also designate the domino or pearl facets as ``terrace
defects'' (Fig.~\ref{fig:terrace-facet}(a)) and calculate their defect
content with a Burgers circuit. Figure~\ref{fig:terrace-facet-defects}
shows some examples of this analysis, where the defect content
relative to the zipper phase was determined. First, we note that the
GB phase junctions are irrelevant to the total defect content of the
terraces since they always appear in opposite pairs that cancel each
other. Next, we sometimes find that a terrace has a finite Burgers
vector content (red circuits in
Fig.~\ref{fig:terrace-facet-defects}). These defects are always
compensated by nearby defects in zipper (yellow extensions to the red
circuits). Therefore, the terraces act as steps in the zipper phase,
with step heights of $n \cdot 37d'_y$. Macroscopically,
Eq.~\ref{eq:pearl-facet-frac} should therefore apply when the pearl
facets are stable.

\section{Conclusion}

The relative stability of GB phases is often investigated for special
or symmetric GBs. Here, we investigate asymmetric tilt GBs as a step
towards understanding general GBs. For our $\Sigma37$c
$[11\overline{1}]$ tilt GBs in Cu, we find different types of defects
in the domino/zipper phase and the pearl phase. The former GB phase
can contain pure steps that can be combined to obtain any inclination
from \ang{0} to \ang{30}, which is the complete range due to the
crystal symmetry. For the pearl phase, different disconnections can be
combined to achieve inclinations up to around \ang{11}. Interestingly,
simulations predict that the combinations of disconnections have lower
defect energies than the pure steps in the domino/zipper phase. This
can be explained by considering that the defect combinations have a
net zero Burgers vector and also act as steps when combined. The
resulting step energy is lower in the pearl phase in the present
case. This leads to a significant reduction of the stability of the
domino phase even at low inclinations compared to the symmetric GB
plane. For inclinations above \ang{11}, faceting of pearl and zipper
phases becomes geometrically necessary and the ratios of the facets
can be predicted.

A comparison to STEM experiments on Cu thin films confirms that the
defect types predicted by the simulations are indeed present in real
samples. Experimentally observed GBs in Al also contain similar
defects, hinting that the present results could be applicable to
different fcc metals. While the defect structures are modeled
correctly, we however observe more domino phase in the experimental Cu
sample than in the simulations. This could indicate that its stability
is underestimated by the interatomic potential. Additionally, a
complex stress state in the thin film sample, which also changes with
temperature during post-deposition annealing, could stabilize the
domino phase. Finally, the experimental sample is much thicker than
the simulation model, slowing down the GB transformation kinetics and
potentially leading to the appearance of residual domino phase.

Overall, these results nevertheless clearly demonstrate that GB
thermodynamics is strongly affected by geometrically necessary
defects, which in turn strongly depend on the atomic structure of the
GB motifs. A full understanding of GB thermodynamics consequently
requires a treatment of GB line defects in addition to the free
energies of the symmetric GB planes.

\section{Acknowledgments}

This project has received funding from the European Research Council
(ERC) under the European Union's Horizon 2020 research and innovation
program (Grant agreement No. 787446; GB-CORRELATE). S.S.\ and G.D.\
acknowledge partial funding from Deutsche Forschungsgemeinschaft (DFG)
within SFB 1394 (project ID 409476157). Y.C.\ was supported by the
National Research Foundation of Korea (NRF) funded by the Korea
government (Ministry of Science and ICT) (No.\ RS-2023-00254343).

S.P.\ conducted most simulations and analyses. L.L.\ performed the
STEM investigations on Cu and S.S.\ performed the STEM investigations
on Al. Y.C.\ ran the structure search with GRIP. R.J., C.H.L.,
G.D., and T.B.\ contributed to the conceptualization of the study,
discussions, and interpretation of the results. C.H.L.\ supervised the
experimental investigations. G.D.\ secured funding via the ERC
advanced Grant GB-CORRELATE and supervised the overall project. T.B.\
performed some analyses, designed the study, and supervised the
simulations. S.P.\ and T.B.\ were responsible for drafting the
manuscript, with all authors contributing to revisions.

\appendix
\section{Supplemental material}

Supplementary material related to this article can be found online
together with this article on arXiv.

\end{document}